\documentclass[aps,prd,showpacs,preprintnumbers,superscriptaddress,nofootinbib]{revtex4}
\usepackage[dvips]{graphicx}
\usepackage{bm,latexsym,amsmath,amssymb,amsfonts}
\newcommand{\stac}[2]{\stackrel{\scriptscriptstyle {#1}}{#2}}
\newcommand*{\cD}{{\cal D}}
\newcommand*{\cF}{{\cal F}}

\newcommand*{\e}{{\rm e}}
\begin{document}

\title{Curvature corrections to the low energy effective theory in 6D
regularized braneworlds}

\author{Tsutomu~Kobayashi}
\email[Email: ]{tsutomu"at"gravity.phys.waseda.ac.jp}
\affiliation{Department of Physics, Waseda University, Okubo 3-4-1, Shinjuku, Tokyo 169-8555, Japan}
\author{Tetsuya~Shiromizu}
\email[Email: ]{shiromizu"at"phys.titech.ac.jp}
\affiliation{Department of Physics, Tokyo Institute of Technology, Tokyo 152-8551,
Japan}
\author{Claudia~de~Rham}
\email[Email: ]{crham"at"perimeterinstitute.ca}
\affiliation{Dept. of Physics \& Astronomy, McMaster University, Hamilton ON, L8S 4M1, Canada\\
Perimeter Institute for Theoretical Physics, 31 Caroline St. N., Waterloo, ON, N2L 2Y5, Canada
}

\begin{abstract}
We study the effective gravitational theory on a brane
in a six-dimensional Einstein-Maxwell model of flux compactification,
regularizing a conical defect as a codimension-one brane.
We employ the gradient expansion technique valid at low energies.
A lowest order analysis showed that standard four-dimensional
Einstein gravity is reproduced on the brane.
We extend this study to include second order corrections in the effective
equations, and show that the correction term is given by a quadratic energy-momentum tensor.
Taking the thin-brane limit where the regularized brane shrinks to the pole,
we find that the second order metric diverges logarithmically on the brane, giving rise
to divergences in the brane effective action. Away from the branes,
the effective action is however well-defined.
\end{abstract}

\preprint{WU-AP/278/08}
\maketitle

\section{Introduction}

Six-dimensional (6D) gravity models with codimension-two branes are intriguing
frameworks in which some of the more interesting physics of extra dimensions,
such as flux stabilization, can occur, while still being sufficiently
simple to be analytically tractable~\cite{cc1, cc-rev, burgess, 6D} (see also
\cite{Hybrid, sugra}).
Models in which the two extra dimensions are large, i.e. sub-millimeter scales, are
interesting because they are not yet ruled out by table-top Newton's
law experiments, but can also resolve the Hierarchy problem through the
so-called ADD mechanism with large extra dimensions~\cite{ADD}.
Furthermore, they have the potential to provide interesting new insights into the
cosmological constant problem~\cite{cc1, cc-rev}, whilst giving clear observational
signatures that will be testable in accelerators, e.g., LHC (Large Hadron
Collider). However, any codimension-two source generically generates
(conical) singularities which makes the study of such systems highly
non-trivial~\cite{Geroch} (see Ref.~\cite{matter}
for recent considerations). Matter sources on codimension-two branes
are therefore to be understood as regularized objects.
In particular, one popular way to regularize a codimension-two brane
is to smooth it out as a codimension-one brane, hence giving it a
finite thickness, and then introduce regular caps to remove the
singularities~\cite{Peloso, Papa}.
This was done explicitly in the model of 6D Einstein-Maxwell flux compactification,
for which it was shown that
4D Einstein gravity is reproduced in the weak field regime~\cite{Peloso, KobaMina}.
Moreover, the low energy effective theory in such a model was derived using the
gradient expansion method~\cite{GE}, also giving rise to conventional
4D gravity at low energies~\cite{FKS}.
The effective theory at lowest order is explicitly finite and independent of
any divergences associated to the codimension-two sources.

In this paper, we extend the gradient expansion analysis to next order to
determine the leading order corrections to standard 4D gravity.
We find that the corrections exhibit clearly the logarithmic divergences
associated with the codimension-two sources. We briefly show
that provided the metric is defined in such a way as to be
finite in the bulk, the effective theory in the bulk remains finite
while diverging logarithmically on the codimension-two brane.

The rest of the paper is organized as follows. In Sec.~II, we carefully
describe the derivation of the low energy effective theory on regularized
branes. In particular we show that the analysis of Ref.~\cite{FKS}
was incomplete but that the main result remains valid. In Sec.~III,
we derive the next order corrections focusing for simplicity on the
case where the regularized branes only contain conformal matter.
In Sec.~IV we discuss the issue associated with the divergence in the thin-brane limit.
Finally we give a summary and discussion in Sec.~V.
We present in Appendix~A some useful computations for the derivation of the effective theory,
and derive the effective action up to second order
in Appendix~B.

\section{Gradient expansion approach: an improved analysis}

The gradient expansion method has been employed
by Fujii, Kobayashi, and Shiromizu (henceforth FKS)~\cite{FKS}  to study
regularized braneworlds in 6D Einstein-Maxwell theory~\cite{Peloso, Papa}.
In that work, the authors showed that standard 4D Einstein gravity is reproduced
at first order in the gradient expansion.
However, not all of the required boundary conditions were satisfied in
that analysis which was therefore incomplete.
In this section, we revisit the FKS gradient expansion approach, showing that
an additional degree of freedom (the lapse function) must be included in the metric ansatz
so that the solution is consistent with all the boundary conditions.
We also show that {\em the main result of FKS remains valid}.

We consider the system described by the action
\begin{eqnarray}
S =\sum_{I=N, 0, S}S_I + \sum_{i=N, S}S_i + \sum_{i=N, S}S_{{\rm GH},i},
\end{eqnarray}
where the bulk action is given by the sum of
\begin{eqnarray}
S_I = \int d^6x\sqrt{-g}\left[\frac{M^4}{2}\left({}^6\! R-\frac{1}{L_I^2}\right)-\frac{1}{4}F_{MN}F^{MN}\right],
\end{eqnarray}
with $I= N$ (for the north cap), $0$ (for the central bulk), and $S$ (for the south cap), and
each of the 4-brane actions $S_i$ is
\begin{eqnarray}
S_i = \int d^5x\sqrt{-q}\left[-\lambda_{(i)}-\frac{v_{(i)}^2}{2}q^{\hat\mu\hat\nu}
\left(\partial_{\hat\mu}\Sigma_{(i)}-\e A_{\hat\mu}\right)
\left(\partial_{\hat\nu}\Sigma_{(i)}-\e A_{\hat\nu}\right)
+{\cal L}_{{\rm m}}^{(i)}\right],
\end{eqnarray}
with $i=N$ (for the north brane) and $S$ (for the south brane).
Capital Latin indices label the 6D coordinates, Greek
indices are restricted to the 4D coordinates, and hatted Greek
indices run over the 5D coordinates along the
branes: $\hat \mu =\mu, \theta$. The Gibbons-Hawking term is given by
\begin{eqnarray}
S_{{\rm GH}, i} =-M^4\int d^5x\sqrt{-q}\big[\big[\hat K\big]\big]_i.\label{gh}
\end{eqnarray}
In the above action $F_{MN}=\partial_MA_N-\partial_NA_M$ is the field strength of
the $U(1)$ gauge field $A_M$, $\lambda_{(i)}$ is the tension of the brane,
$v_{(i)}$ is a constant parameter, $\Sigma_{(i)}$ is a scalar field localized on the brane,
and ${\cal L}_{{\rm m}}^{(i)}$ is the Lagrangian of usual matter.
$\hat K$ is the 5D trace of the extrinsic curvature of the brane and
$[[\cdots]]$ denotes the difference between the two sides of the surface under consideration.
In what follows, we will suppress the labels $I$ and $i$ unless necessary.

Since we are mainly interested in the thin-brane limit of the
effective theory, we focus our analysis on the unwarped case.
Using this implicit assumption, the metric can then be written as
\begin{eqnarray}
ds^2&=&q_{\hat\mu\hat\nu}dx^{\hat\mu}dx^{\hat\nu}+g_{\xi\xi}d\xi^2\notag\\
&=&g_{\mu\nu}(x, \xi)dx^{\mu}dx^{\nu}+L^2
e^{2\zeta(x)\sin^2\! \xi}d\xi^2+L^2B^2e^{2\psi(x, \xi)}\sin^2\!\xi d\theta^2,
\label{metric}
\end{eqnarray}
where the lapse $\zeta(x)$ is the new degree of freedom, and we
assume a cylindrical symmetry for the metric.
Notice that in the gradient expansion, the $(\mu\theta)$ component
is of order ${\cal O} (\varepsilon^{3/2})$,
where $\varepsilon$ is the small expansion parameter
to be specified below. This component is hence suppressed~\cite{FKS}.

The north and south poles are respectively located at $\xi=0$ and $\pi$.
The conical deficit is controlled by the parameters $B$ $(=B_N, B_0, B_S)$.
Since the caps are regular, we require that $B_N=B_S=1$ and $\psi(0)=\psi(\pi)=0$, while in general
we have $B_0\neq 1$.
The regularized branes are located at $\xi = \xi_N$ and $\xi_S$.
Continuity of the induced metric across the brane imposes $L_N = B_0L_0=L_S \;(=: \ell)$.

The system is governed
by the the following set of equations~\cite{FKS}, including the evolution equations,
\begin{eqnarray}
n^\xi\partial_{\xi}K_{\hat\mu}^{\;\hat\nu}+\hat{K}K_{\hat\mu}^{\;\hat\nu}
={}^5\! R_{\hat\mu}^{\;\hat\nu}-\frac{1}{4L^2}\delta_{\hat\mu}^{\;\hat\nu}
-\frac{1}{M^4}\left(F_{\hat\mu M}F^{\hat\nu M}-\frac{1}{8}\delta_{\hat\mu}^{\;\hat\nu}F^2\right)
-{}^5\!D_{\hat\mu}{}^5\!D^{\hat\nu}\zeta\sin^2\!\xi-{}^5\!D_{\hat\mu}\zeta{}^5\!D^{\;\hat\nu}\zeta\sin^4\!\xi,
\label{eveq}
\end{eqnarray}
the Hamiltonian constraint,
\begin{eqnarray}
{}^5\!R+K_{\hat\mu}^{\;\hat\nu}K_{\hat\nu}^{\;\hat\mu}-\hat K^2=\frac{1}{L^2}
-\frac{2}{M^4}\left(F_{\xi M}F^{\xi M}-\frac{1}{4}F^2\right),
\label{hamc}
\end{eqnarray}
and the momentum constraints,
\begin{eqnarray}
{}^5\!D_{\hat\nu}\left(K_{\hat\mu}^{\;\hat\nu}-\delta_{\hat\mu}^{\;\hat\nu}\hat K\right)
=\frac{1}{M^4}F_{\hat\mu M}F^{\xi M}n_\xi,
\label{momc}
\end{eqnarray}
where $K_{\hat\mu}^{\;\hat\nu}$ is the extrinsic curvature of
$\xi=$ constant hypersurfaces, $n^\xi =1/\sqrt{g_{\xi\xi}}$,
$F^2:=F_{MN}F^{MN}$,
and ${}^5\!D_{\hat\mu}$
is the covariant derivative with respect to the 5D metric $q_{\hat\mu\hat\nu}$.
Notice that since the positions of the branes are given by $\xi=$ constant,
the brane extrinsic curvature in~(\ref{gh}) must be identified as the one used in
Eqs.~(\ref{eveq})--(\ref{momc}).
We also have the Maxwell equations
\begin{eqnarray}
\nabla_{N}F^{NM}=0,\label{maxw}
\end{eqnarray}
where $\nabla_N$ is the covariant derivative with respect to the 6D metric.

The evolution equations, constraints, and the Maxwell equations~(\ref{eveq})--(\ref{maxw})
are supplement with the boundary conditions at
the branes and poles. At the poles we will impose the regularity conditions as specified in what follows.
The boundary conditions on the branes are given by the Israel junction conditions,
\begin{eqnarray}
\big[\big[K_{\hat\mu}^{\;\hat\nu}-\delta_{\hat\mu}^{\;\hat\nu}\hat K\big]\big]
=-\frac{1}{M^4}T_{\hat\mu{{\rm (tot)}}}^{\;\hat\nu},
\end{eqnarray}
where
\begin{eqnarray}
T_{\hat\mu{{\rm (tot)}}}^{\;\hat\nu}:=-\lambda\delta_{\hat\mu}^{\;\hat\nu}+
v^2\left[
(\partial_{\hat\mu}\Sigma-\e A_{\hat\mu})(\partial^{\hat\nu}\Sigma-\e A^{\hat\nu})
-\frac{1}{2}
(\partial_{\hat\lambda}\Sigma-\e A_{\hat\lambda})(\partial^{\hat\lambda}\Sigma-\e A^{\hat\lambda})
\delta_{\hat\mu}^{\;\hat\nu}
\right]+T_{\hat\mu}^{\;\hat\nu},
\end{eqnarray}
and $T_{\hat\mu}^{\;\hat\nu}$ comes from the matter Lagrangian ${\cal L}_{{\rm m}}$.
In what follows we assume that there is no matter on the south brane, $T_{\hat\mu}^{\;\hat\nu}|_S=0$.
Since the brane action couples to the gauge field, $F_{MN}$ has a discontinuity at the position of the brane.
This discontinuity is described by the jump conditions
\begin{eqnarray}
\label{JumpMaxwell}
\big[\big[ n^{\xi}F_{\xi\hat\mu}\big]\big]=-\e v^2(\partial_{\hat\mu}\Sigma-
\e A_{\hat\mu}).
\end{eqnarray}

Following FKS, we solve the above set of equations
using the gradient expansion technique.
The metric, extrinsic curvature, and the Maxwell field are expanded as~\cite{FKS}
\begin{eqnarray*}
g_{\mu\nu}=h_{\mu\nu}(x)+\varepsilon g^{(1)}_{\mu\nu}+\cdots,
\quad
\zeta = \zeta^{(0)}+\varepsilon \zeta^{(1)}+\cdots,
\quad
\psi = \psi^{(0)}+\varepsilon \psi^{(1)}+\cdots,
\\
K_{\hat\mu}^{\;\hat\nu}=\stac{(0)}{K_{\hat\mu}^{\;\hat\nu}}
+\varepsilon \!\stac{(1)}{K_{\hat\mu}^{\;\hat\nu}}+\cdots,
\quad
A_{\theta}=A_{\theta}^{(0)}+\varepsilon A_{\theta}^{(1)}+\cdots,\qquad\qquad
\end{eqnarray*}
where the small parameter $\varepsilon$
is the ratio of the bulk curvature scale to the 4D intrinsic curvature scale, $\varepsilon\sim \ell^2|R|$.
Therefore, the covariant derivative with respect to the 4D metric $h_{\mu\nu}$,
${\cal D}_{\mu}$, gives rise to ${\cal O}(\varepsilon^{1/2})$ contributions.

\subsection{Zeroth order result}

The zeroth order evolution equations are given by
\begin{eqnarray}
\frac{1}{L}\partial_{\xi}\!\stac{(0)}{K_{\mu}^{\;\nu}}+
\Big(\stac{(0)}{K_{\lambda}^{\;\lambda}}+\stac{(0)}{K_{\theta}^{\;\theta}}\Big)\stac{(0)}{K_{\mu}^{\;\nu}}
=-\frac{1}{4L^2}+\frac{1}{4}\stac{(0)}{F_{\xi\theta}}\stac{(0)}{F^{\xi\theta}}
,\quad
\frac{1}{L}\partial_{\xi}\!\stac{(0)}{K_{\theta}^{\;\theta}}+
\Big(\stac{(0)}{K_{\lambda}^{\;\lambda}}+\stac{(0)}{K_{\theta}^{\;\theta}}\Big)\stac{(0)}{K_{\theta}^{\;\theta}}
=-\frac{1}{4L^2}-\frac{3}{4}\stac{(0)}{F_{\xi\theta}}\stac{(0)}{F^{\xi\theta}},
\end{eqnarray}
and the Maxwell equation is $\partial_{\xi}(\sin\xi\!\stac{(0)}{F^{\xi\theta}})=0$.
They are solved by
\begin{eqnarray}
\stac{(0)}{K_{\mu}^{\;\nu}}=0,
\quad
\stac{(0)}{K_{\theta}^{\;\theta}}= \frac{1}{L}\cot\xi,
\quad
\stac{(0)}{F_{\xi\theta}}=\ell M^2\sin\xi,
\quad \psi^{(0)}=\zeta^{(0)}=0.
\label{solzero}
\end{eqnarray}
One can easily check that (\ref{solzero}) indeed satisfies the
zeroth order representation of the constraint equations~(\ref{hamc})
and~(\ref{momc}).

The gradient expansion of the brane scalar field is
$\Sigma=\Sigma^{(0)}(\theta, x)+\sigma^{(1)}(x)+\cdots$.
It follows from the equation of motion that $\partial_\theta^2\Sigma^{(0)}=0$,
leading to~\cite{FKS}
\begin{eqnarray}
\Sigma^{(0)}= n\theta+\sigma^{(0)}(x),
\quad n=0, \pm1, ...\,.
\end{eqnarray}
The Israel conditions at the branes are given by
\begin{eqnarray}
\begin{cases}\;
\displaystyle{
\lambda = \frac{v^2}{2}\frac{1}{\ell^2\sin^2\xi }(n-\e A_{\theta}^{(0)})^2}
\\
\; [[ L^{-1} ]]\cot\xi  =- 2\lambda/M^4
\end{cases},
\end{eqnarray}
and the jump condition for the Maxwell field reads
\begin{eqnarray}
~[[ L^{-1} ]]\ell M^2\sin\xi  &=&-\e v^2(n-\e A_{\theta}^{(0)}),
\end{eqnarray}
where $\xi $ should be understood as the position of the brane that we are considering.

\subsection{First order analysis and the recovery of 4D gravity}

The traceless and trace parts of the $(\mu\nu)$ evolution equations at first order
are given by\footnote{Here and hereafter we use ``trace(less)'' in the 4D sense.}
\begin{eqnarray}
\partial_{\xi}\!\stac{(1)}{\mathbb{K_{\mu}^{\;\nu}}}+\cot\xi \stac{(1)}{\mathbb{K_{\mu}^{\;\nu}}} &=&
L\mathbb{R_{\mu}^{\;\nu}},\label{traceless1}
\\
\partial_{\xi}\!\stac{(1)}{K_{\mu}^{\;\mu}}+\cot\xi \stac{(1)}{K_{\mu}^{\;\mu}} &=&
L\left(R+ {\cal F}^{(1)}\right),\label{trace1}
\end{eqnarray}
where $\mathbb{K_{\mu}^{\;\nu}}$ and $\mathbb{R_{\mu}^{\;\nu}}$
are the traceless part of $K_{\mu}^{\;\nu}$ and $R_{\mu}^{\;\nu}$ (the 4D Ricci tensor of $h_{\mu\nu}$),
respectively, $R:=R_{\mu}^{\;\mu}$,
and the first order expansion of the field strength is expressed as
\begin{eqnarray}
M^4{\cal F}^{(1)} :=\stac{(0)}{F_{\xi\theta}}\stac{(1)}{F^{\xi\theta}}
+\stac{(1)}{F_{\xi\theta}}\stac{(0)}{F^{\xi\theta}}.\label{def:F}
\end{eqnarray}
The $(\theta\theta)$ evolution equation is
\begin{eqnarray}
\partial_{\xi}\!\stac{(1)}{K_{\theta}^{\;\theta}}
+2\cot\xi\stac{(1)}{K_{\theta}^{\;\theta}}
+\cot\xi \stac{(1)}{K_{\mu}^{\;\mu}} &=&-\frac{3}{4}
L {\cal F}^{(1)}-\frac{1}{L}\zeta^{(1)},\label{thth1}
\end{eqnarray}
and the
Hamiltonian constraint is
\begin{eqnarray}
2\cot\xi\stac{(1)}{K_{\mu}^{\;\mu}}
=L\left(R+{\cal F}^{(1)}\right).\label{Ham1}
\end{eqnarray}
The momentum constraints are to be discussed below.

The general solution to the traceless evolution equation~(\ref{traceless1}) is given by
\begin{eqnarray}
\stac{(1)}{\mathbb{K}_{\mu}^{\;\nu}}
=-L\mathbb{R}_{\mu}^{\;\nu}\cot\xi +\frac{\chi_{\mu}^{\;\nu}(x)}{\sin\xi},
\end{eqnarray}
where $\chi_{\mu}^{\;\nu}$ is a traceless integration constant. This can be fixed by
imposing the regularity condition at the poles,
\begin{eqnarray}
\stac{(1)}{\mathbb{K}_{\mu}^{\;\nu}}\to 0 \quad(\xi\to 0, \pi),
\end{eqnarray}
and the Israel conditions at the south brane,
\begin{eqnarray}
\Big[\Big[\stac{(1)}{\mathbb{K}_{\mu}^{\;\nu}}\Big]\Big]_S =0.
\end{eqnarray}
We find
\begin{eqnarray}
\stac{(1)}{\mathbb{K}_{\mu}^{\;\nu}}=\mathbb{R_{\mu}^{\;\nu}}\times
\begin{cases}
L_N\tan(\xi/2)\\
\displaystyle{-\frac{\alpha_1}{2}L_0\cot(\xi/2)+\frac{\alpha_2}{2}L_0\tan(\xi/2)}\\
-L_S\cot(\xi/2)
\end{cases},\label{bbK_1sol}
\end{eqnarray}
where we defined
\begin{eqnarray}
\alpha_1&:=&
2 \left( \sin^2(\xi_S/2)+\frac{L_S}{L_0} \cos^2(\xi_S/2)
\right),
\\
\alpha_2&:=&2-\alpha_1= 2\left(1-\frac{L_S}{L_0}\right)\cos^2(\xi_S/2).
\end{eqnarray}
Then, the Israel conditions at the  north brane,
\begin{eqnarray}
\Big[\Big[\stac{(1)}{\mathbb{K}_{\mu}^{\;\nu}}\Big]\Big]_N =-\frac{\mathbb{T}_{\mu}^{\;\nu}}{M^4},
\end{eqnarray}
gives
\begin{eqnarray}
\frac{{\cal V}}{2\pi\ell \sin\xi_N}\mathbb{R_{\mu}^{\;\nu}} =\frac{\mathbb{T_{\mu}^{\;\nu}}}{M^4},
\label{traceless_result}
\end{eqnarray}
where ${\cal V}$ is the volume of the 2D internal space:
\begin{eqnarray}
{\cal V}&:=&2\pi\ell\int^{\pi}_{0} L\sin\xi d\xi
\nonumber\\
\label{calV}
&=&2\pi\ell\sin\xi_N\left[L_N\tan(\xi_N/2)+\frac{1}{2}L_0
\left(\alpha_1\cot(\xi_N/2)-\alpha_2\tan(\xi_N/2)\right)\right]\,.
\end{eqnarray}

The analysis of the trace part equations results in
a complicated expression for a general solution, which is deferred to Appendix~A.
This leads to 15 unspecified functions, which are fixed using the
boundary conditions.
At the poles we require the following regularity conditions:
\begin{eqnarray}
\stac{(1)}{K_{\theta}^{\;\theta}}, \;
\psi^{(1)}, \;A_{\theta}^{(1)} \to0\quad(\xi\to 0, \pi).
\end{eqnarray}
Notice that $\stac{(1)}{K_{\mu}^{\;\mu}}$ is trivially regular at the poles (see Appendix~A).
The continuity of the induced metric implies that $\psi^{(1)}$ is continuous across the  branes.
We also require that $A_{\theta}^{(1)}$ is continuous across the  branes
so that the brane action is well-defined. Thus,
\begin{eqnarray}
\Big[\Big[ \psi^{(1)}\Big]\Big] =0,\quad
\Big[\Big[ A_{\theta}^{(1)}\Big]\Big] =0.
\end{eqnarray}
The Israel conditions and the jump condition for the Maxwell field at the branes
are summarized as follows:
\begin{eqnarray}
(\mu\mu):&&
\Big[\Big[\stac{(1)}{K_{\theta}^{\;\theta}}+\frac{3}{4}\stac{(1)}{K_{\mu}^{\;\mu}}\Big]\Big]
=\frac{v^2}{M^4}\frac{1}{\ell^2\sin^2\xi}\left[
(n-\e A_{\theta}^{(0)})^2\psi^{(1)}+(n-\e A_{\theta}^{(0)})\e A_{\theta}^{(1)}
\right]+\frac{1}{4M^4}T_{\mu}^{\;\mu},
\label{Is_trace}
\\
(\theta\theta):&&
\Big[\Big[ \stac{(1)}{K_{\mu}^{\;\mu}}\Big]\Big]=-\frac{v^2}{M^4}\frac{1}{\ell^2\sin^2\xi}\left[
(n-\e A_{\theta}^{(0)})^2\psi^{(1)}+(n-\e A_{\theta}^{(0)})\e A_{\theta}^{(1)}
\right]+\frac{1}{ M^4}T_{\theta}^{\;\theta},
\label{Is_thth}
\\
(\text{Maxwell}):&&
\Big[\Big[L^{-1}\Big(
\!\stac{(1)}{F_{\xi\theta}}
-\zeta^{(1)}\sin^2\!\xi\!\stac{(0)}{F_{\xi\theta}}
\Big)\Big]\Big]=\e^2v^2 A_{\theta}^{(1)},
\label{Max_jump}
\end{eqnarray}
where we consider the south brane to be empty,
$T_{\mu}^{\;\nu}|_S=T_{\theta}^{\;\theta}|_S=0$.
We are then left with the $3\times2$ regularity conditions, $2\times2$ continuity conditions,
$1\times2$ $(\theta\theta)$ Israel conditions, and $1\times 2$ Maxwell jump conditions.
Finally, considering the trace of the Israel condition on the south brane, we
end up with a total of 15 boundary conditions.
Using them all we can write the $x$-dependent functions
$C_I, \;\zeta^{(1)}_I,\; \Psi_I,\; \Theta_I, \;a_I$
in terms of $R$ and $T_{\theta}^{\;\theta}$.
Finally, the $(\mu\mu)$ Israel condition at the  north brane is used to
derive the trace part of the effective equations.

We have therefore confirmed that all of the boundary conditions can be satisfied consistently.
In fact, this is sufficient for the purpose of deriving the effective equations {\em at first order},
as we can do so without knowing all the integration constants explicitly
(although  this has lead to an incomplete analysis in \cite{FKS} as explained below).

To see this point clearly, it is convenient to use~\cite{FKS}
\begin{eqnarray}
{\cal K}^{(1)}=\stac{(1)}{K_{\theta}^{\;\theta}}+\frac{3}{4}\stac{(1)}{K_{\mu}^{\;\mu}}
+\frac{1}{L}\cot\xi \;\psi^{(1)} +
\frac{L}{M^4}\stac{(0)}{F^{\xi\theta}}\!\!A_{\theta}^{(1)}.
\label{def:K}
\end{eqnarray}
It follows from the evolution equations and Hamiltonian constraint that
\begin{eqnarray}
\partial_{\xi}\left(\sin\xi\;{\cal K}^{(1)}\right)= \frac{1}{4}LR \sin\xi.
\end{eqnarray}
The general solution is given by
\begin{eqnarray}
{\cal K}^{(1)}=-\frac{1}{4}L R \cot \xi +\frac{\chi(x)}{\sin \xi},
\end{eqnarray}
where $\chi$ is an integration constant.

The regularity conditions at the poles read ${\cal K}^{(1)}\to 0$ as $\xi\to0, \pi$.
With the help of the zeroth order junction conditions, Eq.~(\ref{Is_trace}) can be written as
\begin{eqnarray}
\left[\left[{\cal K}^{(1)}\right]\right]_N=\frac{1}{4}\frac{T_{\mu}^{\;\mu}}{M^4},
\quad
\left[\left[{\cal K}^{(1)}\right]\right]_S=0.
\end{eqnarray}
Determining the integration constants by the regularity and
the Israel condition at the  south brane, we obtain
\begin{eqnarray}
{\cal K}^{(1)}= \frac{R}{4}\times
\begin{cases}
L_N\tan(\xi/2)\\
\displaystyle{-\frac{\alpha_1}{2}L_0\cot(\xi/2)+\frac{\alpha_2}{2}L_0\tan(\xi/2)}\\
-L_S\cot(\xi/2)
\end{cases}.\label{bulk_sol_K(1)}
\end{eqnarray}
The Israel condition at the  north brane reduces to
\begin{eqnarray}
\frac{{\cal V}}{2\pi\ell \sin\xi_N}R =-\frac{T_{\mu}^{\;\mu}}{M^4}.
\end{eqnarray}
Combining this with the traceless result~(\ref{traceless_result}), we
finally obtain
the effective equations
\begin{eqnarray}
R_{\mu}^{\;\nu}-\frac{1}{2}\delta_{\mu}^{\;\nu}R = 8\pi G\;\overline{T}_{\mu}^{\;\nu},
\label{eff:Ein}
\end{eqnarray}
where 4D gravitational constant is given by $(8\pi G)^{-1}:=M^4{\cal V}$ and
$\overline{T}_{\mu}^{\;\nu}$ is the energy-momentum tensor
integrated along the $\theta$-direction,
\begin{eqnarray}
\overline{T}_{\mu}^{\;\nu}:=\int T_{\mu}^{\;\nu}\sqrt{g_{\theta\theta}}d\theta = 2\pi\ell \sin\xi_N
T_{\mu}^{\;\nu}.
\end{eqnarray}
Eq.~(\ref{eff:Ein}) shows that standard 4D general relativity is reproduced
on the brane at lowest order.
Notice also that
the momentum constraints,
\begin{eqnarray}
{\cal D}_\nu\! \stac{(1)}{\mathbb{K}^{\;\nu}_{\mu}}-{\cal D}_{\mu}{\cal K}^{(1)}=0,
\end{eqnarray}
are trivially satisfied thanks to the Bianchi identities,
leading to the energy-momentum conservation on the brane, ${\cal D}_{\mu}\overline{T}_{\nu}^{\;\mu}=0$.

In this paper, we will mainly focus on the analysis of the  thin-brane limit
where the regularized branes shrink to the poles: $\xi_N\to 0$ and $\xi_S\to\pi$.
We take the effective energy-momentum tensor
$\overline{T}_{\mu}^{\;\nu}$
to be finite in the limit $\xi_N\to 0$.
The lowest order effective equations~(\ref{eff:Ein}) do not depend explicitly
on the positions of the branes.
They are dependent implicitly on
$\xi_N$ through the volume of the 2D internal space~(\ref{calV}).
This volume remains finite in the limit $\xi_N\to 0$.
(In particular, since the south brane is empty, one can take
$\xi_S=\pi$, in which case the internal space volume simplifies to
${\cal V} = 4 \pi \ell L_0$.)
Therefore, Eq.~(\ref{eff:Ein}) is free from any divergences in the thin-brane limit.

One can compute ${\cal K}^{(1)}$ directly by substituting
the result of Appendix A into the definition of ${\cal K}^{(1)}$ [Eq.~(\ref{def:K})].
We then find that all the integration constants cancel except for $a$,
and one can identify $a=\chi$.
Although the effective equations can be derived by using the special combination of
the variables ${\cal K}^{(1)}$ and only the $(\mu\mu)$ junction conditions,
the remaining boundary conditions~(\ref{Is_thth}) and~(\ref{Max_jump})
must be satisfied by appropriately chosen $C_I$, $\zeta^{(1)}_I$, $\Theta_I$, and $\Psi_I$.
In the FKS analysis~\cite{FKS}, the additional degrees of freedom
$\zeta^{(1)}_I(x)$ were overlooked and the $(\theta\theta)$ Israel junction
condition as well as the Maxwell
jump condition were not consistently satisfied.
We should emphasize that the definition of ${\cal K}^{(1)}$
is precisely the same as that in~\cite{FKS}; the more general metric ansatz~(\ref{metric})
does not give rise to any additional contributions to ${\cal K}^{(1)}$, and hence
FKS have obtained the correct effective equations.

\section{Second order corrections}

We continue to solve the governing equations at second order in the gradient expansion.
Although being straightforward, the general expression for the
second order result can be quite messy. In this section, we therefore focus
on conformally invariant matter, i.e.,
on the traceless energy-momentum tensor for brane matter, $T_{\mu}^{\;\mu}=0$.
We also assume that $T_{\theta}^{\;\theta}=0$.
Using the lowest order effective equations, we see that the scalar curvature vanishes,
$R=0$, and so all the ``trace part'' integration constants
also vanish. Thus, we have
\begin{eqnarray*}
\stac{(1)}{K_{\mu}^{\;\mu}}\,=\,
\stac{(1)}{K_{\theta}^{\;\theta}}
\,=\,\psi^{(1)}=\zeta^{(1)}=\stac{(1)}{F_{\xi\theta}}
=A_{\theta}^{(1)}=0.
\end{eqnarray*}
This restriction greatly simplifies the analysis while capturing the
main features of the thin-brane limit.

Integrating $\stac{(1)}{\mathbb{K}_{\mu}^{\;\nu}}$, we get
\begin{eqnarray}
g_{\mu\nu}^{(1)}(x, \xi)=2 U (\xi)\mathbb{R}_{\mu\nu},
\end{eqnarray}
where
\begin{eqnarray}
U(\xi)=
\begin{cases}
\displaystyle{
-2L_N^2\ln\left[
\frac{\cos(\xi/2)}{\cos(\xi_N/2)}
\right]
}
\qquad\qquad\qquad\qquad\qquad\qquad(\xi<\xi_N)
\\
\displaystyle{
- \alpha_1L_0^2
\ln\left[\frac{\sin(\xi/2)}{\sin(\xi_N/2)}\right]
- \alpha_2 L_0^2
\ln\left[
\frac{\cos(\xi/2)}{\cos(\xi_N/2)}\right]
}
\;\;(\xi_N<\xi<\xi_S)
\\
\displaystyle{
-2L_S^2 \ln\left[ \frac{\sin(\xi/2)}{\sin(\xi_S/2)}\right]
- \tilde \alpha L_0^2
}
\qquad\qquad\qquad\qquad\quad(\xi>\xi_S)
\end{cases},\label{defu}
\end{eqnarray}
with
\begin{eqnarray}
\tilde \alpha:=\alpha_1
\ln\left[\frac{\sin(\xi_S/2)}{\sin(\xi_N/2)}\right]
+
\alpha_2\ln\left[
\frac{\cos(\xi_S/2)}{\cos(\xi_N/2)}\right].
\end{eqnarray}
The integration constants are determined so that $g^{(1)}_{\mu\nu}(x, \xi_N)=0$
(i.e., so that the brane induced metric is given by $h_{\mu\nu}$)
and
the metric is continuous across each of the branes.
It is instructive to summarize here the properties of the function $U(\xi)$:
\begin{eqnarray}
&&\qquad\partial_\xi^2U+\cot\xi \partial_{\xi}U=L^2,\label{u1}
\\
&&\qquad \partial_\xi U(0) = \partial_\xi U(\pi)=0,\label{u2}
\\
&& [[U]]_{N, S}=0,\;U(\xi_N)=0,\;[[\partial_{\xi}U/L]]_S=0.\label{u3}
\end{eqnarray}


The second order part of the 5D Ricci tensor is given by
$\left[{}^5\! R_{\theta}^{\;\theta}\right]^{(2)}=0$ and
\begin{eqnarray}
\left[{}^5\! R_{\mu}^{\;\nu}\right]^{(2)}=-2 U (\xi)
{\cal S}_{\mu}^{\;\nu}(x)\,,
\label{5R2nd}
\end{eqnarray}
where
\begin{eqnarray}
{\cal S}_{\mu}^{\;\nu}(x):= \mathbb{R}_{\mu\lambda}\mathbb{R}^{\lambda\nu}
-\frac{1}{2}\cD_{\lambda}\cD_{\mu}\mathbb{R}^{\lambda \nu}-
\frac{1}{2}\cD^{\lambda}\cD^{\nu}\mathbb{R}_{\mu\lambda}
+\frac{1}{2} \cD^2\mathbb{R}_{\mu}^{\;\nu}.
\end{eqnarray}

The evolution equations at second order reduce to
\begin{eqnarray}
\frac{1}{L}\Big(\partial_\xi\! \stac{(2)}{K_{\mu}^{\;\nu}}+\cot\xi\stac{(2)}{K_{\mu}^{\;\nu}}\Big)
&=&\left[{}^5\! R_{\mu}^{\;\nu}\right]^{(2)}+\frac{1}{4}\cF^{(2)}\delta_{\mu}^{\;\nu},
\label{munu2}
\\
\frac{1}{L}\Big(\partial_\xi\! \stac{(2)}{K_{\theta}^{\;\theta}}+2\cot\xi\stac{(2)}{K_{\theta}^{\;\theta}}
+\cot\xi\stac{(2)}{K_{\mu}^{\;\mu}}\Big)
&=& -\frac{3}{4}\cF^{(2)}-\frac{1}{L}\zeta^{(2)},
\end{eqnarray}
and the Hamiltonian constraint is
\begin{eqnarray}
L\left(\left[{}^5\! R_{\mu}^{\;\mu} \right]^{(2)}+\cF^{(2)}\right)
=2\cot\xi\stac{(2)}{K_{\mu}^{\;\mu}}-L \stac{(1)}{\mathbb{K}_{\mu}^{\;\nu}}
\stac{(1)}{\mathbb{K}_{\nu}^{\;\mu}},\label{ham2}
\end{eqnarray}
where we defined
\begin{eqnarray}
M^4\cF^{(2)}:=\stac{(0)}{F_{\xi\theta}}\stac{(2)}{F^{\xi\theta}}
+\stac{(2)}{F_{\xi\theta}}\stac{(0)}{F^{\xi\theta}}.
\end{eqnarray}

Let us first consider the traceless part of Eq.~(\ref{munu2}).
We can immediately integrate the traceless evolution equation to obtain
\begin{eqnarray}
\stac{(2)}{\mathbb{K}_{\mu}^{\;\nu}}&=&
4 L_N^3\tilde {\cal S}_{\mu}^{\;\nu}\left\{
\frac{\cos\xi-1}{2\sin\xi}-\frac{\ln[\cos(\xi/2)\cos(\xi_N/2)]}{\sin\xi}-\cot\xi\ln\left[
\frac{\cos(\xi/2)}{\cos(\xi_N/2)}
\right]
\right\}
\qquad\qquad(\xi<\xi_N),
\\
\stac{(2)}{\mathbb{K}_{\mu}^{\;\nu}}&=&
\frac{2L_0^3\alpha_1}{\sin\xi} \tilde {\cal S}_{\mu}^{\;\nu}\left\{
\frac{\cos\xi}{2}+\ln\left[\sin(\xi/2)\right]-\cos\xi\ln\left[
\frac{\sin(\xi/2)}{\sin(\xi_N/2)}
\right]\right\}
\nonumber\\&&\qquad\qquad
+\frac{2L_0^3\alpha_2}{\sin\xi}\tilde {\cal S}_{\mu}^{\;\nu}\left\{
\frac{\cos\xi}{2}-\ln\left[\cos(\xi/2)\right]-\cos\xi\ln\left[
\frac{\cos(\xi/2)}{\cos(\xi_N/2)}
\right]
\right\}
+\frac{\Xi _{\mu}^{\;\nu}(x)}{\sin\xi}
\quad (\xi_N<\xi<\xi_S),
\end{eqnarray}
and
\begin{eqnarray}
\stac{(2)}{\mathbb{K}_{\mu}^{\;\nu}}&=&
4 L_S^3\tilde {\cal S}_{\mu}^{\;\nu}\left\{
\frac{\cos\xi+1}{2\sin\xi}+\frac{\ln[\sin(\xi/2)\sin(\xi_S/2)]}{\sin\xi}-\cot\xi\ln\left[
\frac{\sin(\xi/2)}{\sin(\xi_S/2)}
\right]
\right\}
\nonumber\\&&\qquad\qquad
-2\tilde \alpha L_SL_0^2\tilde{\cal S}_{\mu}^{\;\nu} \frac{1+\cos\xi}{\sin\xi}
\qquad\qquad\qquad\qquad\qquad\qquad\qquad\qquad\qquad(\xi>\xi_S),
\end{eqnarray}
where $\tilde{\cal S}_{\mu}^{\;\nu}$ is the traceless part of ${\cal
S}_{\mu}^{\;\nu}$: $\tilde{\cal S}_{\mu}^{\;\nu}:={\cal
S}_{\mu}^{\;\nu}-(1/4)\delta_{\mu}^{\;\nu}{\cal
S}_{\lambda}^{\;\lambda}$.
Notice that $\cD_{\nu}\tilde{\cal S}_{\mu}^{\;\nu}=0$ when $R=0$.
The integration constants in the capped regions are determined using the regularity conditions:
$\stac{(2)}{\mathbb{K}_{\mu}^{\;\nu}}\to0$ as $\xi\to0, \pi$.
There is then one remaining integration constant $\Xi_{\mu}^{\;\nu}$ in the bulk region,
which is to be fixed by imposing the Israel condition at the  south brane,
\begin{eqnarray}
\Big[\Big[\stac{(2)}{\mathbb{K}_{\mu}^{\;\nu}}\Big]\Big]_S =0.
\end{eqnarray}

The traceless part of the Israel conditions at the  north brane is given by
\begin{eqnarray}
\frac{{\cal V}}{2\pi\ell \sin\xi_N}\mathbb{R}_{\mu}^{\;\nu}
-\Big[\Big[ \stac{(2)}{\mathbb{K}_{\mu}^{\;\nu}}\Big]\Big]_N=\frac{1}{M^4}\mathbb{T}_{\mu}^{\;\nu},
\end{eqnarray}
which, with some manipulation, reduces to
\begin{eqnarray}
\frac{{\cal V}}{2\pi\ell \sin\xi_N}\mathbb{R}_{\mu}^{\;\nu}
+\frac{\beta}{\sin\xi_N} \tilde{\cal S}_{\mu}^{\;\nu}=\frac{1}{M^4}\mathbb{T}_{\mu}^{\;\nu},
\label{pre_eff2}
\end{eqnarray}
where
\begin{eqnarray}
\beta&:=&-2\int_0^\pi U(\xi)L\sin\xi d\xi
\nonumber\\&=&
-2\left(1-\cos\xi_N+4\ln[\cos(\xi_N/2)]\right)L_N^3
-2\left(1+\cos\xi_S+4\ln[\sin(\xi_S/2)]\right)L_S^3
\nonumber\\&&
+2(1+\cos\xi_S)\tilde \alpha L_SL_0^2
-(\cos\xi_N-\cos\xi_S)(\alpha_1+\alpha_2)L_0^3
\nonumber\\&&\quad
+2\left\{
(1-\cos\xi_S)\ln\left[\frac{\sin(\xi_S/2)}{\sin(\xi_N/2)}\right]\alpha_1
-(1+\cos\xi_S)\ln \left[\frac{\cos(\xi_S/2)}{\cos(\xi_N/2)}\right]\alpha_2
\right\}L_0^3.
\end{eqnarray}
Using the leading order effective equations we can write $\tilde{\cal S}_{\mu}^{\;\nu}$
in terms of the integrated energy-momentum tensor $\overline{\mathbb{T}}_{\mu}^{\;\nu}$.
Thus, we obtain the effective equations
\begin{eqnarray}
\mathbb{R}_{\mu}^{\;\nu}=\frac{\overline{\mathbb{T}}_{\mu}^{\;\nu}}{M_{{\rm Pl}}^2}
+8\pi \ell \beta\frac{M^4}{M_{{\rm
Pl}}^6}\tilde\Pi_{\mu}^{\;\nu}\,,
\label{effctive2nd}
\end{eqnarray}
where
\begin{eqnarray}
\tilde\Pi_{\mu}^{\;\nu}=
-\frac{1}{4}
\overline{\mathbb T}_{\mu\lambda} \overline{\mathbb T}^{\lambda\nu}
+\frac{1}{16}\delta_{\mu}^{\;\nu}
\overline{\mathbb T}_{\rho\lambda} \overline{\mathbb T}^{\rho\lambda }
+\frac{1}{8M_{\rm pl}^2}\left[{\cD}^\lambda {\cD}_\mu \overline{\mathbb T}^{\;\nu}_\lambda
+{\cD}_\lambda {\cD}^\nu \overline{\mathbb T}_\mu^{\;\lambda}
-{\cD}^2 \overline{\mathbb T}_\mu^{\;\nu}
\right].
\end{eqnarray}

For completeness let us look at the second order extension of the quantity defined in Eq.~(\ref{def:K}):
\begin{eqnarray}
{\cal K}^{(2)}:=\stac{(2)}{K^{\;\theta}_\theta}+\frac{3}{4}
\stac{(2)}{K^{\;\mu}_\mu}+ \frac{1}{L}\cot\xi\;\psi^{(2)}
+\frac{L}{M^4}\stac{(0)}{F^{\xi \theta}}\!\!A_\theta^{(2)}.
\end{eqnarray}
The trace part of the second order evolution equations~(\ref{munu2}) and the Hamiltonian constraint~(\ref{ham2})
give the evolution equation for ${\cal K}^{(2)}$:
\begin{eqnarray}
\frac{1}{\sin\xi}\partial_\xi \left(\sin \xi \,{\cal K}^{(2)} \right)
&=&\frac{1}{4} L[\, {}^5\!R^{\;\mu}_\mu]^{(2)}
-\frac{1}{2}L  \stac{(1)}{\mathbb{K}_{\mu}^{\;\nu}}\stac{(1)}{\mathbb{K}_{\nu}^{\;\mu}}
\nonumber\\
&=&
-\frac{1}{2} \left[
U/L^2+(\partial_{\xi}U/L^2)^2
\right]L^3\mathbb{R}_{\mu}^{\;\nu}\mathbb{R}_{\nu}^{\;\mu}\,,
\label{ev:K:2nd}
\end{eqnarray}
while the trace part of the Israel conditions reduces to
\begin{eqnarray}
[[{\cal K}^{(2)}]]_{N, S}=0.
\end{eqnarray}
We also require the regularity at the poles: ${\cal K}^{(2)}\to 0$ as $\xi\to0, \pi$.
The solution to Eq.~(\ref{ev:K:2nd}) is written in terms of three integration constants,
which are fixed by the regularity at the poles and the Israel condition at the  south brane.
The resulting solution then automatically satisfies the Israel condition
at the  north brane.
Using Eq.~(\ref{u1}) one finds that the solution can be expressed simply as
\begin{eqnarray}
{\cal K}^{(2)}=-\frac{1}{2L} U\partial_{\xi}U\,\mathbb{R}_{\mu}^{\;\nu}\mathbb{R}_{\nu}^{\;\mu}.
\label{sol2K}
\end{eqnarray}
Eqs.~(\ref{u2}) and~(\ref{u3}) guarantee that the solution satisfies the required
boundary conditions. Since ${\cal K}^{(2)}(\xi_N)=0$ and $[[{\cal K}^{(2)}]]_N=0$, the trace part
of the effective equations is trivial at second order.

The momentum constraints at second order reduce to
\begin{eqnarray}
{\cal D}_\nu\stac{(2)}{ \mathbb{K}_{\mu}^{\;\nu}}
-\stac{(1)}{\Gamma_{\mu\nu}^{\lambda}}\stac{(1)}{\mathbb{K}_{\lambda}^{\;\nu}}
-{\cal D}_\nu {\cal K}^{(2)}=0,\label{mom2}
\end{eqnarray}
where
$
\stac{(1)}{\Gamma_{\mu\nu}^{\lambda}}\,=\frac{1}{2}h^{\lambda\sigma}
\left(
\cD_\mu g^{(1)}_{\nu\sigma}+\cD_\nu g^{(1)}_{\mu\sigma}-\cD_\sigma g^{(1)}_{\mu\nu}
\right)
$
and $g^{(1)}_{\mu\nu}$ is traceless.
Recalling that $\cD_{\nu}\tilde{{\cal S}}_{\mu}^{\;\nu}=0$,
one can easily check that Eq.~(\ref{mom2}) is consistently satisfied.

\subsection*{An example: the radiation-dominated universe}

Using the effective equations~(\ref{effctive2nd}), we can obtain
the modified Friedmann equation in the radiation-dominated universe as
\begin{eqnarray}
H^2 = \frac{\rho}{3M_{{\rm Pl}}^2}\left(1+\frac{\rho}{\rho_*}\right),
\end{eqnarray}
where $H$ is the Hubble parameter, $\rho$ is the energy density and
\begin{eqnarray}
\rho_*:=\frac{3M_{{\rm Pl}}^2}{8\pi}\frac{{\cal V}}{\ell \beta}.
\end{eqnarray}
This Friedmann equation is valid when $\rho\ll |\rho_*|$.
The $\rho^2$ correction term here is similar to what has been found
in the Randall-Sundrum brane cosmology~\cite{RS}.

\section{4D Effective action and thin brane limit}

\subsection{Effective action}

In this section we present the 4D low energy effective
action to first and second order. The derivation of these effective
theories is presented in detail in Appendix B and here we focus on the
main results.

Using the first order expression for the metric in the bulk, one can
integrate the action over the 2D internal space and obtain the
following 4D action when including the boundary
contributions,
\begin{eqnarray}
S_{{\rm 4D}}^{(1)}=\frac{{\cal V}M^4}{2}\int d^4x\sqrt{-h}\,R
+\int d^4x \sqrt{-h}\,\overline{{\cal L}}_{{\rm m}},
\end{eqnarray}
where for the matter action we use the dimensional reduction over the $\theta$-direction
\begin{eqnarray}
\int d^5x\sqrt{-q}\,{\cal L}_{{\rm m}}
\to\int d^4x \sqrt{-h}\,\overline{{\cal L}}_{{\rm m}}\,.
\end{eqnarray}

Proceeding similarly for the second order terms, we obtain the
second order effective action
\begin{eqnarray}
\label{2ndaction}
S_{{\rm 4D}}^{(2)}
=2\pi\ell M^4\int d^4x \sqrt{-h} \;\frac{\beta}{4}\mathbb{R_{\mu}^{\;\nu}}\mathbb{R_{\nu}^{\;\mu}},
\end{eqnarray}
where we assumed conformal matter on the north brane and an empty south
brane.

\subsection{Thin-brane limit}

Since no matter is present on the south brane, the limit
$\xi_S\rightarrow \pi$ is regular and can be taken
without further ado. (In this limit, we simply have $\alpha_1=2$ and $\alpha_2=0$.)
On the north brane, on the other hand,
the limit $\xi_N\rightarrow 0$ should be taken with care as it leads
to a divergent term:
\begin{eqnarray}
\beta\to-4L_0^3 -8L_0^3\ln[\sin(\xi_N/2)].
\end{eqnarray}
At the linearized level,
we expect these divergences to be renormalizable so that any
brane and bulk observables are finite~\cite{Goldberger, deRham}.
More precisely, the propagator of fields living in the bulk are finite
when evaluated away from the brane, while the propagator of fields
confined to the brane (matter fields) are finite despite their
coupling with bulk fields.

In the present case, the coefficient of the nonlinear term $\tilde{\cal S}_{\mu}^{\;\nu}$
(or $\tilde\Pi_{\mu}^{\;\nu}$)
diverges and the effective equations are likely to be invalid in the thin-brane limit.
Furthermore, the metric is not finite away from
the brane in this limit [see Eq.~(\ref{defu})].
We now investigate whether it is possible somehow to avoid these divergences
by reconsidering the choice of the integration constant
(which is attributed to the different boundary condition imposed at the north brane).
For this purpose we instead consider the modified metric
defined as
\begin{eqnarray}
\hat g^{(1)}_{\mu\nu}(x,\xi)=g^{(1)}_{\mu\nu}(x,\xi)-4 L_0^2 \mathbb{R}_{\mu\nu}
\ln \left[\sin (\xi_N/2)\right]\,,
\end{eqnarray}
so as to remove any divergences in the bulk.
The additional term here corresponds to the change of the integration constant.
The first order correction to the bulk metric is then
expressed as
\begin{eqnarray}
\hat g^{(1)}_{\mu\nu}(x,\xi)= -4 \mathbb{R}_{\mu\nu}\times
\begin{cases}
\displaystyle{
  L_N^2 \ln \left[\frac {\cos (\xi/2)}{\cos (\xi_N/2)}\right]+
   L_0^2 \ln \left[\sin \xi_N/2\right] \quad (0<\xi<\xi_N)}
\\
\displaystyle{
  L_0^2 \ln \left[\sin(\xi/2)\right] \qquad\qquad\qquad\qquad\qquad  ( \xi_N<\xi<\pi)}
\end{cases}
.
\end{eqnarray}
The metric evaluated on the brane
is now given by
\begin{eqnarray}
\hat h_{\mu\nu}:=h_{\mu\nu}-4L_0^2\mathbb{R}_{\mu\nu}\ln[\sin(\xi_N/2)],
\end{eqnarray}
rather than $h_{\mu\nu}$. We observe that in this representation,
the metric in the bulk is independent of the regularization
procedure and taking the thin-brane limit poses no problem. On the
brane, the metric diverges logarithmically where
the regularizing scale $\xi_N$ is sent to zero, as expected for
codimension-two systems~\cite{Geroch}.

In terms of this new metric, the second order
effective theory is still given by~(\ref{2ndaction}), but with
$\beta$ now replaced by
\begin{eqnarray}
\hat \beta&:=&\beta+ 4L_0^2\frac{\cal V}{2\pi \ell} \ln \left[\sin
(\xi_N/2)\right]
\nonumber\\
&=&-4L_0^3\bigg\{
\cos^2(\xi_N/2)+\frac{L_N^3}{L_0^3}\sin^2(\xi_N/2)+2\frac{L_N^3}{L_0^3}\ln[\cos(\xi_N/2)]
+2\left(1-\frac{L_N}{L_0}\right)\sin^2(\xi_N/2)\ln[\sin(\xi_N/2)]
\bigg\}.
\end{eqnarray}
Notice that $h_{\mu\nu}$ is not the induced metric on the brane, but only
its finite part, in this notation. Actually, $h_{\mu\nu}$ is the metric on the south
pole: $g_{\mu\nu}(x, \pi)=h_{\mu\nu}+\hat g_{\mu\nu}^{(1)}(x, \pi)=h_{\mu\nu}$.
The effective theory will therefore be a good approximation
for observers away from the brane, where the metric remains finite.
In the thin brane limit $\xi_N\to 0$, $\hat \beta$
remains finite: $\hat \beta\to-4L_0^3$, so that the effective theory for $h_{\mu\nu}$ (away from the brane)
is well-defined.
In order to see whether this theory makes sense for branes observers, one should
study how couplings of brane matter fields with gravity ought to be renormalized
and give finite physical observables. This is however beyond
the scope of this study.

Before closing this section, it is worth mentioning that
for a relativistic particle living on the brane,
the quadratic energy-momentum tensor $\tilde\Pi_{\mu}^{\;\nu}$ vanishes.
In this case the thin-brane limit $\xi_N\to0$ is manifestly regular at second order.
This fact (at least partly) explains why the procedure to construct shockwave solutions
in codimension-two braneworlds~\cite{matter, wave} works so well.

\section{Summary and discussion}

In this paper we have derived the higher order corrections
to the effective theory of 6D Einstein-Maxwell theory, regularizing the conical branes as
codimension-one objects.
We first improved the previous analysis of
\cite{FKS} at lowest order and confirmed the validity of their central result,
hence recovering 4D Einstein gravity as an effective theory.
We then derived the next order correction, focusing on conformally
invariant matter for simplicity, and analyzed the thin-brane limit in which the regularized brane
shrinks to the pole.
At lowest order the 4D effective action is free from any
divergences, but we have found that at second order the brane metric diverges in the thin-brane limit.
We can instead define the effective theory with respect to the metric that remains
finite in the bulk. With the metric defined as such, the effective action is well-behaved even at
second order. As expected, however, this metric diverges on the brane and
one should carefully treat the brane couplings before making any
physical conclusions. In particular, we expect such couplings to be
renormalized as in \cite{Goldberger, deRham}.
This issue is left for future studies.

\acknowledgments

TK is supported by the JSPS under Contract No.~19-4199. TS is
supported by Grant-Aid for Scientific Research from Ministry of
Education, Science, Sports and Culture of Japan (Nos.~17740136,
17340075, and 19GS0219), the Japan-U.K., Japan-France and
Japan-India Research Cooperative Programs. CdR wishes to thank the
Tokyo Institute of Technology for its
hospitality while part of this work was being completed.
Research at McMaster is supported by the Natural
Sciences and Engineering Research Council of Canada.
Research at Perimeter Institute for Theoretical Physics is supported
in part by the Government of Canada through NSERC and by the
Province of Ontario through MRI.


\appendix

\section{Solving the trace part equations at first order}\label{app:trace}

In this Appendix we shall solve the ``trace part'' evolution equations.
Below we will introduce several integration constants without stating so.

Combining the trace equation~(\ref{trace1}) and Hamiltonian constraint~(\ref{Ham1}) we find
\begin{eqnarray}
\partial_{\xi}\!\stac{(1)}{K_{\mu}^{\;\mu}}-\cot\xi \stac{(1)}{K_{\mu}^{\;\mu}}=0,
\label{app:ev:mumu}
\end{eqnarray}
which is solved to give
\begin{eqnarray}
\stac{(1)}{K_{\mu}^{\;\mu}} =L C(x)\sin\xi.
\end{eqnarray}
Now the $(\theta\theta)$ evolution equation reduces to
\begin{eqnarray}
\partial_{\xi}\!\stac{(1)}{K_{\theta}^{\;\theta}}+2\cot\xi\stac{(1)}{K_{\theta}^{\;\theta}}
+\frac{5}{2}LC \cos\xi = \frac{3}{4}LR-\frac{2}{L}\zeta^{(1)},
\end{eqnarray}
and the general solution is given by
\begin{eqnarray}
\stac{(1)}{K_{\theta}^{\;\theta}}=
\left(\frac{3}{16}LR-\frac{1}{4L}\zeta^{(1)}\right)\frac{2\xi-\sin(2\xi)}{\sin^2\!\xi}
-\frac{5}{6}LC\sin\xi+\frac{\Theta(x)}{\sin^2\!\xi}.
\end{eqnarray}
Note that $\stac{(1)}{K_{\theta}^{\;\theta}}$ is given
in terms of the metric functions as
\begin{eqnarray}
\stac{(1)}{K_{\theta}^{\;\theta}}=\frac{1}{L}\partial_{\xi}\psi^{(1)}-\frac{1}{L}\zeta^{(1)}\cos\xi\sin\xi.
\end{eqnarray}
This leads to
\begin{eqnarray}
\psi^{(1)}=-\left(\frac{3}{8}L^2R-\frac{\zeta^{(1)}}{2}\right)\xi \cot \xi
+\frac{5}{6}L^2C  \cot \xi -L \Theta \cot \xi +\frac{1}{2}\zeta^{(1)} \sin^2 \xi+\Psi(x).
\end{eqnarray}
Lowering the indices  of $\stac{(1)}{F^{\xi\theta}}$ in~(\ref{def:F}) we get
\begin{eqnarray}
{\cal F}^{(1)}=\frac{2}{L^2\ell M^2}\frac{1}{\sin\xi}\stac{(1)}{F_{\xi\theta}}-\frac{2}{L^2}\left(
\psi^{(1)}+\zeta^{(1)}\sin^2\!\xi\right).\label{F_to_F**}
\end{eqnarray}
This and the Hamiltonian constraint,
$
{\cal F}^{(1)}=-R+2C\cos\xi,
$
yield
\begin{eqnarray}
\stac{(1)}{F_{\xi\theta}}= \partial_{\xi}A_{\theta}^{(1)}=\ell M^2\sin\xi\left[
-\frac{1}{2}L^2 R+\psi^{(1)}+\zeta^{(1)}\sin^2\!\xi +L^2C \cos\xi
\right].\label{app:F=}
\end{eqnarray}
Integrating~(\ref{app:F=}) we obtain
\begin{eqnarray}
A_{\theta}^{(1)}/\ell M^2 & = &
\frac{L^2R}{2} \cot \xi -\frac{11}{24}L^2C \cos (2\xi)
-\left(\frac{3}{8}L^2R-\frac{\zeta^{(1)}}{2}\right)(\xi \sin \xi+\cos \xi) \nonumber \\
& & -L  \Theta  \sin \xi -\Psi  \cos \xi
+\frac{3}{2}\zeta^{(1)} \left(-\cos \xi +\frac{1}{3} \cos^3 \xi \right)+La(x).
\end{eqnarray}
In the above we have 15 unspecified quantities:
\begin{eqnarray}
C_I, \;\zeta^{(1)}_I,\; \Psi_I,\; \Theta_I, \;a_I
\quad (I= N, 0, S).
\end{eqnarray}

\section{Four-dimensional Effective Action}
\label{App:EeffAct}

In this Appendix, we derive the effective action
and confirm that the effective equations are deduced from the action.

We start with the first order bulk action.
Using the six-dimensional Einstein equation, ${}^6\!R=\frac{3}{4}L^{-2}+\frac{1}{4}M^{-4}F^2$,
and the Hamiltonian constraint~(\ref{Ham1}),
we have
\begin{eqnarray}
\int d^6x\sqrt{-g}\left[\frac{M^4}{2}\left({}^6\! R-\frac{1}{L_I^2}\right)-\frac{1}{4}F^2\right]
&=&M^4\int d^6x\sqrt{-g}\left(
\frac{1}{4L^2}-\frac{1}{8M^4}F^2
\right)
\nonumber\\
&=&2\pi\ell M^4\int d\xi d^4x\sqrt{-h}\sin\xi
\left(\frac{1}{4}LR-\frac{1}{2}\cot\xi\stac{(1)}{K_\mu^{\;\mu}}\right).
\end{eqnarray}
Using the evolution equation~(\ref{app:ev:mumu}) we find
\begin{eqnarray}
\int d\xi\cos\xi \stac{(1)}{K_\mu^{\;\mu}} &=&\left[\sin\xi \stac{(1)}{K_\mu^{\;\mu}}\right]_0^\pi
-\int d\xi\sin\xi\partial_\xi\!\stac{(1)}{K_\mu^{\;\mu}}
\nonumber\\
&=&
-\sum_{i=N,S}\sin\xi_i \Big[\Big[\stac{(1)}{K_\mu^{\;\mu}}\Big]\Big]_i
-\int d\xi\cos\xi \stac{(1)}{K_\mu^{\;\mu}}.
\end{eqnarray}
Therefore, the bulk action reduces to
\begin{eqnarray}
\frac{M^4{\cal V}}{4}\int d^4x\sqrt{-h}R
+2\pi\ell M^4\sum_{i=N,S}
\int d^4x\sqrt{-h}
\frac{1}{4}\sin\xi_i \Big[\Big[\stac{(1)}{K_\mu^{\;\mu}}\Big]\Big]_i.
\end{eqnarray}

Each of the brane actions and surface contributions is given by
\begin{eqnarray}
-M^4\int d^5x\sqrt{-q}\,[[\hat{K}]]+\int d^5x\sqrt{-q}\left[-\lambda-\frac{v^2}{2}
(\partial_{\hat\mu}\Sigma-\e A_{\hat\mu})(\partial^{\hat\mu}\Sigma-\e A^{\hat\mu})
+{\cal L}_{{\rm m}}\right].
\end{eqnarray}
Up to first order, this therefore reduces to
\begin{eqnarray}
-2\pi\ell M^4\int d^4x\sqrt{-h} \sin\xi_i \left(\left[\left[{\cal K}^{(1)}\right]\right]+\frac{1}{4}\Big[\Big[
\stac{(1)}{K_{\mu}^{\;\mu}}\Big]\Big]\right),
\end{eqnarray}
where we used the zeroth order jump condition for the Maxwell field~(\ref{JumpMaxwell}).
At the south brane we have $[[{\cal K}^{(1)}]]_S=0$, while at the
north brane, the bulk solution~(\ref{bulk_sol_K(1)}) implies
$[[{\cal K}^{(1)}]]_N=-\frac{1}{4}R\cdot{\cal V}/2\pi\ell
\sin\xi_N$.
Therefore, the sum of the bulk, branes, and surface term contributions
up to first order is
\begin{eqnarray}
S_{{\rm 4D}}^{(1)}=\frac{{\cal V}M^4}{2}\int d^4x\sqrt{-h}\,R
+\int d^4x \sqrt{-h}\,\overline{{\cal L}}_{{\rm m}},
\end{eqnarray}
where a reduction
\begin{eqnarray}
\int d^5x\sqrt{-q}\,{\cal L}_{{\rm m}}
\to\int d^4x \sqrt{-h}\,\overline{{\cal L}}_{{\rm m}}
\end{eqnarray}
is understood.

We now compute the second order of this action.
We focus on the case where the matter energy-momentum tensor is traceless
and hence use the bulk solution presented in the previous section.
The bulk action at second order is given by
\begin{eqnarray}
2\pi\ell M^4\int d^4xd\xi\sqrt{-h}\sin\xi\left(-\frac{1}{4}L{\cal F}^{(2)}\right)
=2\pi\ell M^4\int d^4xd\xi\sqrt{-h}\sin\xi\left(\frac{1}{4}L[R_{\mu}^{\;\mu}]^{(2)}
-\frac{1}{2}\cot\xi\stac{(2)}{K_{\mu}^{\;\mu}}+\frac{1}{4}L
\stac{(1)}{\mathbb{K}_{\mu}^{\;\nu}}\stac{(1)}{\mathbb{K}_{\nu}^{\;\mu}}
\right),\label{bulk2action}
\end{eqnarray}
using the Hamiltonian constraint at second order~(\ref{ham2}).
Similarly we find that
\begin{eqnarray*}
\int d\xi \cos\xi\stac{(2)}{K_{\mu}^{\;\mu}}&=&\left[\sin\xi \stac{(2)}{K_{\mu}^{\;\mu}}\right]^{\pi}_{0}
-\int d\xi \sin\xi\partial_{\xi}\!\stac{(2)}{K_{\mu}^{\;\mu}}
\\
&=&
\left[\sin\xi \stac{(2)}{K_{\mu}^{\;\mu}}\right]^{\pi}_{0}
-\int d\xi\left(\cos\xi\stac{(2)}{K_{\mu}^{\;\mu}}-\sin\xi L\mathbb{K_{\mu}^{\;\nu}}\mathbb{K_{\nu}^{\;\mu}}
\right),
\end{eqnarray*}
and the bulk action~(\ref{bulk2action}) then reduces to
\begin{eqnarray}
-2\pi\ell M^4\int d^4x d\xi\sqrt{-h}\frac{1}{2}U(\xi)L\sin\xi{\cal S}_\mu^{\;\mu}
+2\pi\ell M^4\sum_{i=N,S}
\int d^4x\sqrt{-h}
\frac{1}{4}\sin\xi_i \Big[\Big[\stac{(2)}{K_\mu^{\;\mu}}\Big]\Big]_i.\label{bulkaction2}
\end{eqnarray}
The second order brane actions and the Gibbons-Hawking terms are expressed as
\begin{eqnarray}
-2\pi\ell M^4\sum_{i=N, S}\int d^4x\sqrt{-h}\sin\xi_i\left(
\left[\left[{\cal K}^{(2)}\right]\right]_i + \frac{1}{4}\Big[\Big[
\stac{(2)}{K_{\mu}^{\;\mu}} \Big]\Big]_i
\right)=-2\pi\ell M^4\sum_{i=N, S}\int d^4x\sqrt{-h}\;\frac{1}{4}\sin\xi_i
 \Big[\Big[ \stac{(2)}{K_{\mu}^{\;\mu}}\Big]\Big],\label{brane2ac}
\end{eqnarray}
using the bulk solution~(\ref{sol2K}) to eliminate $[[{\cal K}^{(2)}]]$.
Notice that Eq.~(\ref{brane2ac}) cancels the second term in
Eq.~(\ref{bulkaction2}), and hence we end up with
\begin{eqnarray}
S_{{\rm 4D}}^{(2)}
=2\pi\ell M^4\int d^4x \sqrt{-h} \;\frac{\beta}{4}\mathbb{R_{\mu}^{\;\nu}}\mathbb{R_{\nu}^{\;\mu}}.
\end{eqnarray}




\end{document}